\documentclass[prl,floatfix,twocolumn,aps,showpacs,superscriptaddress]{revtex4-1}
\usepackage{graphicx}
\usepackage{amsmath}
\usepackage{amssymb}
\usepackage{natbib}
\usepackage{color}

\begin{document}
\newcommand{\beq}{\begin{equation}}
\newcommand{\eeq}{\end{equation}}

\title{Rigidity Loss in Disordered Systems: Three Scenarios}

\author{Wouter G. Ellenbroek}
\affiliation{Department of Applied Physics and Institute for Complex Molecular Systems, Eindhoven University of Technology, Postbus 513, NL-5600 MB Eindhoven, The Netherlands}
\author{Varda F. Hagh}
\affiliation{Department of Physics, Arizona State University, Tempe, AZ 85287-1504, USA}
\author{Avishek Kumar}
\affiliation{Department of Physics, Arizona State University, Tempe, AZ 85287-1504, USA}
\author{M. F. Thorpe}
\affiliation{Department of Physics, Arizona State University, Tempe, AZ 85287-1504, USA}
\affiliation{Rudolf Peierls Centre for Theoretical Physics, University of Oxford, 1 Keble Rd, Oxford OX1 3NP, England}
\author{Martin van Hecke}
\affiliation{Huygens-Kamerlingh Onnes Lab, Universiteit Leiden, P.O.~Box~9504, NL-2300 RA Leiden, The Netherlands}
\affiliation{FOM Institute AMOLF, Science Park 104, 1098 XG Amsterdam, The Netherlands}

\date{\today}

\begin{abstract}

We reveal significant qualitative differences in the rigidity transition of
three types of disordered network materials: randomly diluted spring networks,
jammed sphere packings, and
\emph{stress-relieved} networks that are diluted using a protocol that
avoids the appearance of floppy regions.
The marginal
state of jammed and stress-relieved networks are globally
isostatic, while marginal randomly diluted
networks show both overconstrained and underconstrained regions.
When a single bond is added to or removed from these isostatic
systems, jammed networks become globally overconstrained or floppy,
whereas the effect on stress-relieved networks is more local and limited.
These differences are also reflected in the linear elastic properties and point
to the highly effective and unusual role of global self-organization in
jammed sphere packings.
\end{abstract}


\maketitle

Disordered elastic networks and sphere packings represent a large class of
amorphous athermal materials, ranging from (bio)polymer networks to granular
media and foams~\cite{BroederszRMP2014,alexander,durianbubble}.
Random networks of springs lose their rigidity when enough springs are cut;
this random bond dilution process is known as rigidity percolation
(RP)~\cite{Feng1984,Feng1985,Sahimi1998,Jacobs1995,Jacobs1996}. Packings of
soft spheres do the same when their confining pressure is lowered towards zero:
this is called
(un)jamming~\cite{jamnature,OHern2003,mvhreview,liunagelreview,carlzilla}.
These rigidity loss scenarios have been studied extensively, in particular for
the simplest cases of networks of harmonic springs~\cite{Jacobs1995,Jacobs1996}
or soft frictionless harmonic
spheres~\cite{OHern2003,mvhreview,liunagelreview,carlzilla}. In that case, the
linear elastic properties of packings can be mapped to those of a spring
network, where each contact is replaced by the appropriate
spring~\cite{wyartthesis,wyart08,ellenbroekEPL09}.  Lowering the pressure, the
number of bonds in the equivalent network decreases.

Given this close correspondence, it is surprising that the nature of the RP and unjamming transitions, and of their respective marginally rigid states, are significantly different.
For packings of a large number ($N$) of soft spheres, extensive studies have shown that \emph{(i)}  the connectivity, i.e., the average number of contacts $z$
per particle, goes to $z_c=2D+\mathcal{O}(1/N)$ at the marginal point, where $D$ is
the space dimension~\cite{jamnature,OHern2003,mvhreview,liunagelreview,carlzilla,
durianbubble,Moukarzel1998,Tkachenko1999,DagoisBohy2012,Goodrich2012}, \emph{(ii)}
the system remains homogeneously jammed up to the point of unjamming (with the exception of individual loose
particles called rattlers or very rare small particle clusters)~\cite{OHern2003}, and \emph{(iii)} the shear modulus, $G$ vanishes as $\Delta z := z-z_c$ whereas
the bulk modulus $K$ remains finite when $\Delta z \rightarrow 0$~\cite{jamnature,OHern2003,mvhreview,liunagelreview,carlzilla,wyartthesis}.
In contrast, in rigidity percolation of generic networks, extensive studies have revealed that for large systems
\emph{(i)} the connectivity $z$, which gives the average number of springs per
node, approaches $z_c=3.9612\ldots < 2D$ for the bond diluted triangular
network~\cite{Jacobs1995,Jacobs1996}, \emph{(ii)} the largest rigid cluster takes on a
heterogeneous, fractal shape, and \emph{(iii)} both the shear modulus, $G$, and
bulk modulus, $K$ smoothly vanish at the critical point in a way typical for a
second order phase transition~\cite{Jacobs1995,Jacobs1996}.

To understand these differences, we note that the small difference in
$z_c$ points to a huge, qualitative difference between jammed and random
networks. Based on extensions of the ideas of Maxwell~\cite{Maxwell1864}), a
simple mean field argument locates the marginal point where the number of
degrees of freedom ($DN$ coordinates) is balanced by the number of constraints
($zN/2$ bonds) at $z=2D$. This argument is exact if all the constraints are
independent and there is a single rigid cluster. If there are redundant bonds, $z_c$ can deviate from $2D$,
although proper counting of \emph{actual} degrees of freedom and
\emph{independent} constraints would remove this apparent violation of
Maxwell's criterion~\cite{Calladine1978}. Indeed, the rigid network in RP
contains both redundant constraints (bonds) and flexible hinges (sites)
at the marginal point so that $z_c\ne 2D$. In contrast, we will show that
sphere packings at the jamming transition are
isostatic everywhere: nothing can move (except a few
rattlers) and \emph{every bond is essential} for the rigidity of the network.
Jammed systems show a high
degree of organization, leading to highly non-generic networks~\cite{ellenbroekEPL09}.

\begin{figure}[!tb]
\includegraphics[width=5cm]{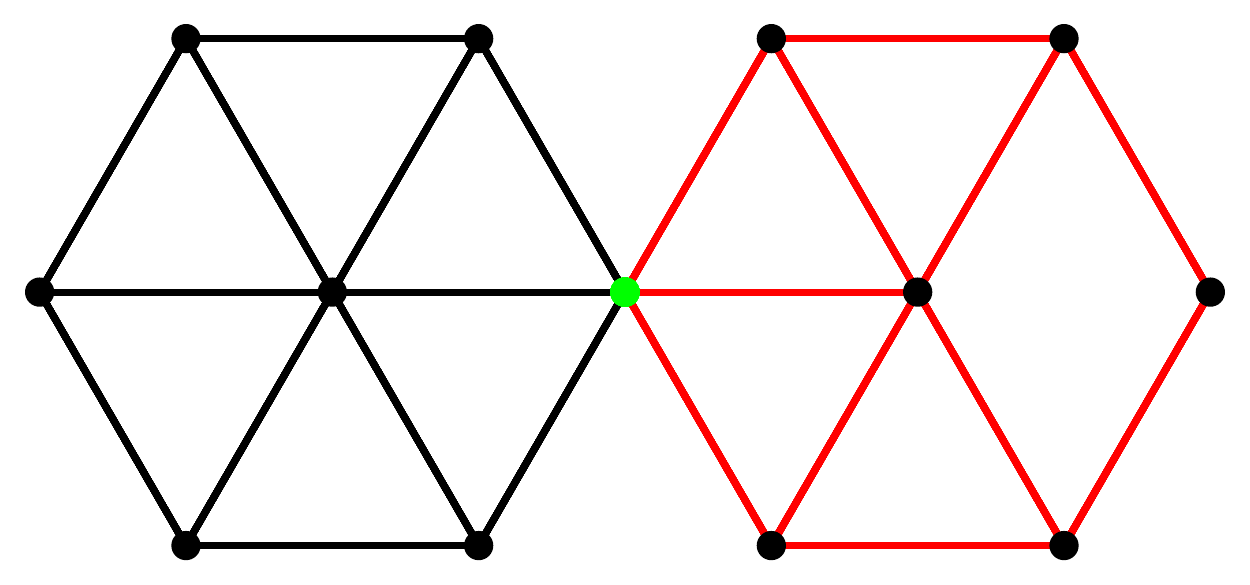}
\caption{Rigid region decomposition, where there are two rigid regions, one (black bonds) overconstrained and the other isostatic (red bonds), separated by a hinge (light green site). The sites which are not hinges are colored black.}
\label{fig:teach}
\end{figure}

Several open questions thus arise: What is different in the topology and
geometry of the underlying networks of random springs and jammed packings? Can
we conceive other families of networks with different rigidity loss
transitions? Here we address these questions, by determining
the overconstrained and underconstrained regions using the pebble game~\cite{Jacobs1995,Jacobs1996}.
This is an integer algorithm that analyzes the topology of generic spring
networks, by a very effective decomposition of such networks into rigid
regions, with both unstressed (isostatic) and stressed (overconstrained or
superfluous~\cite{Thorpe2000}) rigid regions, and the hinges that separate
rigid regions.

Figure~\ref{fig:teach} illustrates such an analysis for a small network. The 12
black bonds (Fig.~\ref{fig:teach} left) might carry finite forces whilst
maintaining force balance: such bonds are redundant, as any one of these bonds
could be removed and the remainder would still be rigid, and are called
\emph{stressed}.  We emphasize that a stressed bond typically, but not
necessarily, carries a finite force: the concept of stressed/redundant bonds
should not be confused with, e.g.,  the
prestress~\cite{Wyart2005PRE,ellenbroekEPL09}.  The 11 red bonds
(Fig.~\ref{fig:teach} right) show a rigid cluster that is exactly isostatic,
and removal of any of these bonds would break the cluster.  Such bonds are
called unstressed, and necessarily carry zero force.  Finally, the green node
in the center of this network is a hinge (defined as a site that belongs to at
least two rigid clusters). For more complex networks,  the pebble game is an
effective algorithm to unambiguously determine the rigid
clusters~\cite{Jacobs1995,Jacobs1996}.

\begin{figure}[!tb]
\includegraphics[width=\columnwidth]{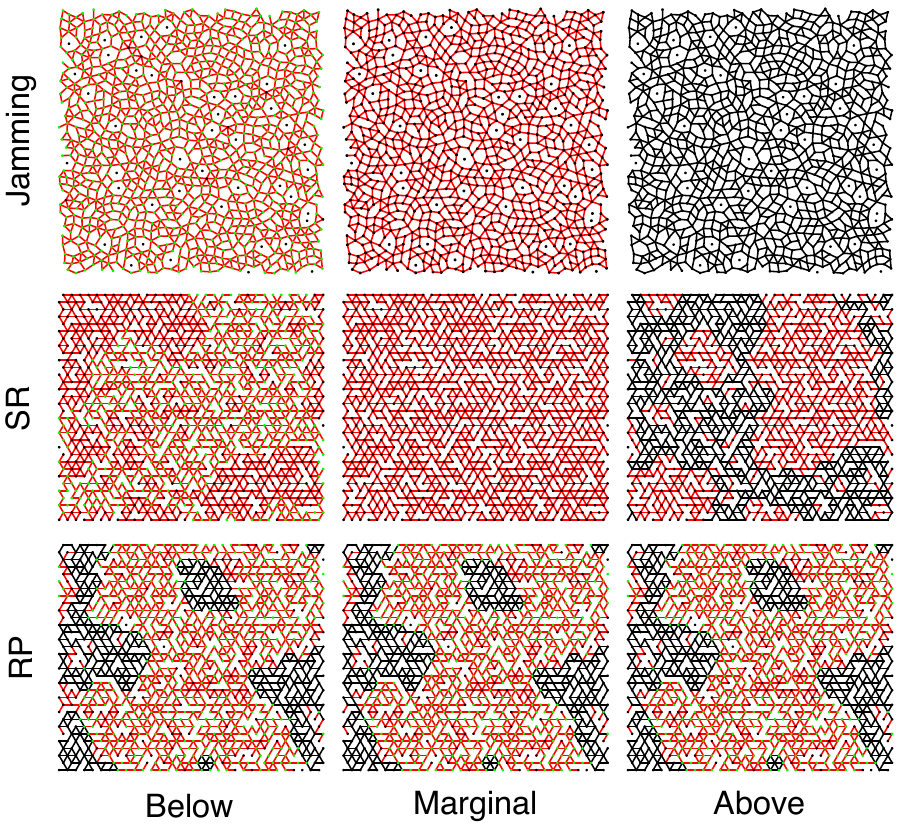}
\caption{Pebble game results for a jammed packing (top row), a stress-relieved triangular network (middle row)
and rigidity percolation (bottom row).
The center panel is the marginal case in all three panels, with the left panel having a single bond removed and the right panel a single bond restored.
The marginal states of both jammed systems as well as the SR network is fully isostatic (red), whereas the
marginal state for RP features floppy modes (involving the light green hinge sites) and has
34\% of all bonds stressed (black).}
\label{fig:networks}
\end{figure}

\emph{Pebble game analysis.---}We will now characterize three families of
network topologies by the pebble game.  Unless otherwise stated RP will refer
to the \emph{bond diluted triangular network} in this Letter, which is the best
studied system. For all networks, we use periodic (wrap-around) boundary conditions.

Figure~\ref{fig:networks} shows dramatic differences in the nature of the marginal
states depending on the physical process that generates these networks.
The top row shows a jammed-packing-derived network at the marginal state
(center), one contact above it (right) and one contact below (left), obtained
by randomly removing bonds from a very weakly jammed packing ($z<4.01$).
Strikingly, in the marginal state of the jammed network, all bonds are isostatic (red),
just above it, the whole system is overconstrained (black), and when a
single bond is removed, almost every site becomes a hinge (green). In terms of
the network topology, this is a massively first order transition.
In the bottom row of Fig.~2, the gentle evolution through the marginal state in RP is shown. The
marginal state contains both isostatic and redundant pieces in the percolating
rigid backbone, as well as significant numbers of green hinges --- adding or
removing a single bond hardly changes the configuration, typical of a
second order transition.

We now
introduce a third family of networks that becomes isostatic everywhere at
their marginal point --- as in jamming --- by cutting bonds randomly, but only if they are stressed.
This stress-relieving (SR) cutting algorithm leads, by construction, to the percolating
marginally rigid cluster being \emph{precisely and exactly isostatic}
everywhere, without any overconstrained or underconstrained regions. This
also means that in both jamming and SR (but not RP) the transition happens at the mean field Maxwell point, so that the mean coordination is $2D$ with zero redundant constraints anywhere.

In the middle row of Fig.~\ref{fig:networks} we show the pebble game analysis for SR
cutting, starting from a triangular network. An isostatic state with a single
cluster is produced at the marginal point, reminiscent of the jammed state.
However, this marginal state is very different in character: both adding or removing a bond has a less dramatic effect than in jamming. Hence,
isostaticity everywhere is not the only nontrivial feature of the jammed state:
its organization is such that its globally isostatic state is changed
\emph{everywhere} by the addition or subtraction of a \emph{single constraint}, in
stark contrast to SR networks.

\begin{figure}[tb]
\includegraphics[width=0.9\columnwidth]{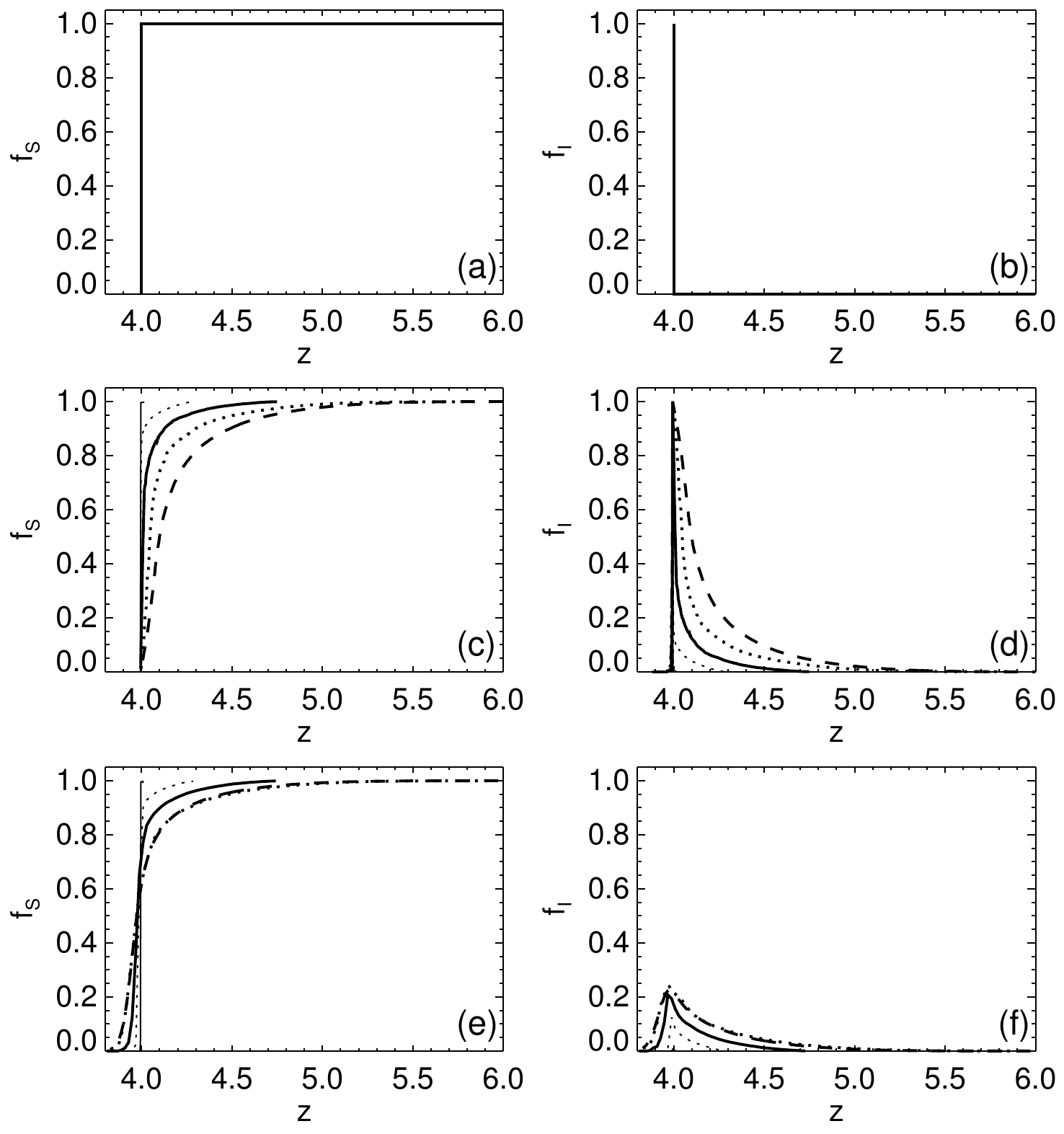}
\caption{Fraction of stressed (left) and  isostatic (right)
bonds in the rigid backbone for jamming (top), stressed bond dilution (middle)
and random bond dilution (bottom). In c-f, line styles indicate the starting
point for bond removal: jammed networks at $z_j=4.01$ (solid, thin),
$z_j=4.3$ (dotted, thin), $z_j=4.7$ (solid, thick),
$z_j=5.98$ (dotted, thick) and triangular (dashed, thick).
Data are averaged over 300 triangular nets or
25--50 jamming-derived networks.}
\label{fig:fractions}
\end{figure}

Both stressed and random bond removal can be performed
on any initial configuration, including jamming-derived networks at given
connectivity $z_j$. Doing so yields \emph{two} two-parameter families of networks,
each characterized by $z$ \emph{and} $z_j$.
Starting with $z_j$ close to $2D$, we can for example probe how,
and how quickly, the network topology crosses over from jammed to generic or SR-like.

In Fig.~\ref{fig:fractions} we compare the fractions of stressed and isostatic bonds
for jamming (top row), SR (middle row) and RP (bottom row), where the latter two
have initial configurations corresponding to jammed networks at four different values of $z$ or a
triangular net. For jamming, the fraction of
stressed bonds, $f_s$, discontinuously jumps from one to zero, and the
fraction of isostatic bonds, $f_i$, jumps from zero to one when $z$ is
lowered, consistent with the picture shown in Fig.~\ref{fig:networks}.
This happens because in jammed sphere packings only contacts that carry a
positive force can be detected and therefore all
bonds in the network must be stressed.
For \emph{random} bond dilution, $f_s(z)$ and $f_i(z)$ remain continuous
irrespective of $z_j$,
and for large $z_j$, these functions smoothly approach those of the triangular
net.

In the middle row of Fig.~\ref{fig:fractions} we show $f_s(z)$ and $f_i(z)$
for the same five families of networks  for \emph{stressed} bond dilution.
The data shown here appear to have a discontinuity around $z=4$;
it is an open question whether this discontinuity persists in the thermodynamic
limit.  For $z_j=5.98$, the apparent jump is small, and the curves are
closer to those of the triangular net. However, we still see
deviations from the triangular case which is surprising given that here we have
to cut almost 1/3 of the bonds to reach the critical point.
 For smaller $z_j$, the apparent jumps in
$f_s$ and $f_i$ grow, approaching the step functions of jamming --- this is
easy to understand, as for $z_j \rightarrow 4$ an increasingly small fraction
of bonds gets removed before reaching $z=4$.

\emph{Discontinuous response to bond addition and removal.---}The response to addition or removal of bonds is a measure for the
degree of organization in the network, and to quantify the discontinuous response
at the marginal point more precisely, we introduce two new indices. The
first is $h$, the ease-of-breakup index which is defined by removing one bond randomly from
the marginal state, counting the number of new green hinges, averaging
over every bond in the network, and
dividing by the number of sites so that
$0 < h < 1$. The second is $s$, the ease-of-stressing index, defined by
adding one bond randomly, counting the number of
new stressed bonds, averaging over all bonds and dividing by the number of bonds so that
$0 < s < 1$.
High values of $h$ and $s$ imply strong self-organization of the network.

We find that in networks representing packings near unjamming the index $h \approx 0.97$ and $s
\approx 0.98$ (cf. top row of Fig.~\ref{fig:networks}),
while for RP networks, both indices are very small ($h \approx 0.0003$ and $s \approx 0.001$) as expected for a second
order transition (see Fig.~\ref{fig:networks}). Intermediate values of $h$ and $s$ are
found for SR
($h \approx 0.28 \pm 0.04$ and $s \approx 0.47 \pm 0.05$), where the spread is specific to our
system sizes and is expected to go down for larger systems.
We have made an additional isostatic marginal state by adding bonds to an empty triangular net, avoiding adding stressed bonds, which
also produces a marginal isostatic state, but with even lower index values:
$h \approx 0.21 $ and $s\approx 0.40$.
The large values of both $h$ and $s$ for the jammed state show how remarkably
self-organized it is.

To understand the large $h$ index for jamming, we start from the globally
isostatic jammed network at the critical point: according to Laman's
theorem~\cite{laman70}, the number of bonds equals $2N-3$ and the number of
bonds $b$ in subgraphs of $n$ nodes satisfies $b \le 2n-3$.  After we remove a
bond, only subgraphs that have precisely $2n-3$ bonds are isostatic.  Examples
of these are $n\!=\!3$ triangles or $n\!=\!4$ double triangles (Fig.~2). Here
all nodes are at the cluster's edge and are hinges; ``black dots'' can only
arise in the interior of isostatic clusters.  The large value of $h$ thus
implies that $n>4$ isostatic clusters are rare in jamming, compared to SR and RP.

We now suggest that large $n$ isostatic clusters are suppressed due to the
homogeneity of jammed systems, using a variation on a well-known bond cutting
argument~\cite{Wyart2005EPL,Goodrich2013SM,mvhreview,liunagelreview}. Consider
a large (hypothetical) isostatic cluster $C$ with $n$ nodes and $2n-3$ \emph{internal}
connections, and $n_e$ nodes at the edge of $C$.  All ${\cal O}(n_e)$
connections that cross the boundary of $C$ (for SR and RP there may be 
fewer) do not contribute to internal connections, so that the mean contact
number of $C$ is $4+\mathcal{O}(n_e/n)$ --- as $n_e \sim \sqrt{n}$, this is
significantly above the global mean contact number $4$, even for relatively
large clusters (for a $n=100$ circular cluster we estimate $z \approx 4.3$).
Whereas RP and SR systems below the marginal point clearly have such subgraphs,
these become extremely unlikely for jammed systems. Thus, the $h$ index
in jamming is much larger than in SR or RP because spatial fluctuations in local contact numbers are
smaller~\cite{Henkes2010}.  How precisely this
homogeneity arises remains an open problem.

To understand the large $s$ index for jamming, we note that for jammed
networks all bonds carry a positive force and are stressed, as jammed systems
are at finite pressure. For SR and RP networks there is no positivity condition
on the contact forces, and both isostatic zero force regions and stressed
regions where positive and negative forces precisely balance can occur. This
difference is clearly illustrated  in SR and RP networks above the marginal
point, where stressed regions can have convex edges where forces of opposite
sign balance --- this is ruled out in jamming. We believe that such differences
also underlie the inequality of the $s$ index for jamming and SR.

\begin{figure}[tb]
\includegraphics[width=\columnwidth]{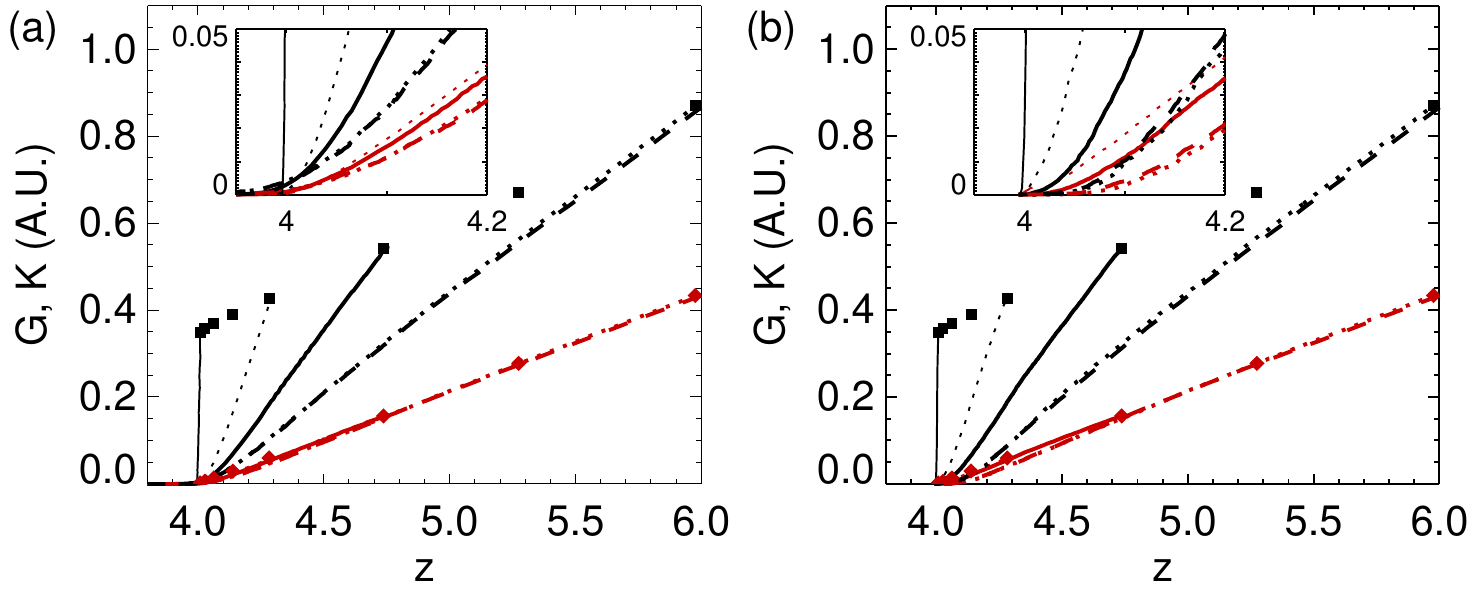}
\caption{Shear modulus $G$ (red) and bulk modulus $K$ (black) for (a) random bond dilution and
(b) stressed bond dilution.
As in Fig.~\ref{fig:fractions}, the initial condition is the network of a jammed packing at
$z_j=4.01$ (solid, thin),
$z_j=4.3$ (dotted, thin), $z_j=4.7$ (solid, thick),
$z_j=5.98$ (dotted, thick) and triangular networks (dashed, thick). as the initial
condition. Insets show zoom-ins around the transition. Solid squares and diamonds denote the moduli of the jammed
packings as published earlier in Ref.~\cite{ellenbroekEPL09}.}
\label{fig:moduli}
\end{figure}

\emph{Elastic moduli.---}We calculate the elastic moduli of the networks in linear
response from the dynamical matrix~\cite{leonforte,ellenbroekPRL06,ellenbroekPRE09}. In Fig.~\ref{fig:moduli} we show
shear ($G$) and bulk ($K$) moduli as a function of $z$ for the same four values of $z_j$ as in Fig.~3
and for the generic triangular net, both for random bond dilution and for stressed-bond-only dilution.
Clearly, a very simple scenario unfolds:
\emph{(1)} For $z_j \approx 6$, the functions $G(z)$ and $K(z)$ are virtually identical to those for
bond dilution of triangular nets. \emph{(2)} $G(z)$ is essentially independent of $z_j$,
consistent with our earlier observations~\cite{ellenbroekEPL09}.
\emph{(3)} The behavior of $K$ is richer. For jammed networks with $z=z_j$, $K$ weakly depends on $z$
but remains finite [$K_j(z=4)>0$]. However, for all $z_j$ that we have investigated, we find that upon bond dilution $K$ vanishes as
\begin{equation}
K(z,z_j)=K_j(z_j) \left[(z_j-z)/(z_j-z_c)\right]^\alpha~,
\end{equation}
where $\alpha$ is close to unity. Our systems are too small to precisely
determine $\alpha$, although the smoothing near $z=4$ is consistent with
$\alpha \approx 1.4$ as found for 2D triangular nets.

Is this difference in moduli related to $h$ and $s$? Strictly speaking, no:
it is the network's geometry, not topology, which determines the elastic response 
(even small geometric perturbations of networks, be they
quasicrystals~\cite{Stenull2014} or jammed~\cite{bastiaanflorijnunpublished},
can strongly perturb $K$). 
However, both the the large value of $s$ and the finite value of $K$,  
are intimately connected to the repulsive nature of contacts in jamming~\cite{wyartthesis,ellenbroekEPL09,Stenull2014}.
Clearly the network reorganizations of jammed systems when they are decompressed
(such geometric reorganizations are absent in SR and RP), leads
to  networks where finite positive contact forces can balance, and $h$ and $s$ tend to one.

\emph{Discussion.---}It was known that jammed networks had to satisfy the Maxwell condition globally and had to satisfy the Hilbert criterion locally~\cite{alexander}, but neither of those imply the self-organization in terms of rigid cluster analysis that we uncover.
From a design perspective, our two-parameter families of networks are
attractive because they allow to independently set the ratio $G/K$ of elastic moduli and the connectivity $z$ (Fig.~\ref{fig:moduli}).
Fully random networks are non-optimal in propagating rigidity, as unhelpful stressed regions remain in the backbone. SR networks are better, but still become soft against compression at their marginal point.  Jamming can be seen as a strategy to find special, perhaps optimal geometries of spring networks in terms of propagating rigidity and resistance to compression, although jammed networks are not the only ones that have finite $K$ at the marginal point~\cite{Stenull2014}.
We have not been able to come up with algorithms that generate networks
with the same intricate network topologies as jamming, and suggest that whether this
is possible remains an important open problem~\cite{Yan2014,Lopez2013}.

Finally, many other marginal networks  have been studied recently~\cite{broedersz11,Sun2012,Mao2013}. Square and kagome
lattices with randomly added braces, which are even more homogeneous than jammed networks, were shown to also have a very sharp
rigidity transition~\cite{zhangchenmao2014preprint} with (in our terminology) $h$ and $s$ close to 1, consistent with our findings.
One alternative protocol to create networks that are isostatic everywhere
was introduced by Lopez \emph{et al.}~\cite{Lopez2013}.
For small $N$, these networks become macroscopically floppy upon removal of a single
bond, but this effect disappears as $N$
increases, and we expect that their networks are similar
to our SR networks, with
$K\to0$. Another recent conditional cutting protocol allows for the independent
tuning of the ratio of bulk and shear moduli~\cite{carlandrea2015preprint}.
We hope that our work will inspire work to analyze such network topologies, leading to
better understanding which other families of networks can be constructed, with distinct properties of the stressed and isostatic bonds, hinges, $h$ and $s$ indices, and elastic moduli.

\begin{acknowledgments}
We acknowledge discussions with N. Upadhyaya and V. Vitelli, who did early calculations on
the bulk modulus in a two-parameter family of networks.
WGE acknowledges support from NWO/VENI. MvH acknowledges support from NWO/VICI.
The work at Arizona Sate University was supported by the National Science
Foundation  under grant DMR 0703973NSF. AK would like to acknowledge funding
from GAANN P200A090123 and the ARCS Foundation.
\end{acknowledgments}

\end{document}